\documentclass[aps,twocolumn,superscriptaddress,showpacs]{revtex4} 
\usepackage{epsf,latexsym} 
\usepackage{times}

\newcommand\N{{\mathrm {I\!N}}} 

\newcommand{\pr}{Phys. Rev.}
\newcommand{\bra}[1]{\langle #1 |} 

\newcommand{\ket}[1]{| #1 \rangle}

\newcommand{\ra}{{\rightarrow}}

\newcommand{\be}{\begin{equation}}
\newcommand{\ee}{\end{equation}} 
\newcommand{\ignore}[1]{}
 
\def\bbbone{{\mathchoice {\rm 1\mskip-4mu l} {\rm 1\mskip-4mu l} 
{\rm 1\mskip-4.5mu l} {\rm 1\mskip-5mu l}}} 

\def\bbbc{{\mathchoice {\setbox0=\hbox{$\displaystyle\rm C$}\hbox{\hbox 
to0pt{\kern0.4\wd0\vrule height0.9\ht0\hss}\box0}} 
{\setbox0=\hbox{$\textstyle\rm C$}\hbox{\hbox 
to0pt{\kern0.4\wd0\vrule height0.9\ht0\hss}\box0}} 
{\setbox0=\hbox{$\scriptstyle\rm C$}\hbox{\hbox 
to0pt{\kern0.4\wd0\vrule height0.9\ht0\hss}\box0}} 
{\setbox0=\hbox{$\scriptscriptstyle\rm C$}\hbox{\hbox 
to0pt{\kern0.4\wd0\vrule height0.9\ht0\hss}\box0}}}} 

\def\bbbz{{\mathchoice {\hbox{$\sf\textstyle Z\kern-0.4em Z$}} 
{\hbox{$\sf\textstyle Z\kern-0.4em Z$}} 
{\hbox{$\sf\scriptstyle Z\kern-0.3em Z$}} 
{\hbox{$\sf\scriptscriptstyle Z\kern-0.2em Z$}}}} 
 
\newcommand{\putfig}[2]{$$\leavevmode\hbox{\epsfxsize=#2 cm 
\epsffile{#1.eps}}$$}

\begin{document}

\title{Testing Bell's Inequality with Ballistic Electrons in Semiconductors}
\author{Radu Ionicioiu}
\author{Paolo Zanardi}
\affiliation{Istituto Nazionale per la Fisica della Materia (INFM)}
\affiliation{Institute for Scientific Interchange (ISI), Villa Gualino, 
Viale Settimio Severo 65, I-10133 Torino, Italy}

\author{Fausto Rossi}
\affiliation{Istituto Nazionale per la Fisica della Materia (INFM)}
\affiliation{Institute for Scientific Interchange (ISI), Villa Gualino, 
Viale Settimio Severo 65, I-10133 Torino, Italy}
\affiliation{Dipartimento di Fisica, Politecnico di Torino, 
Corso Duca degli Abruzzi 24, 10129 Torino, Italy}

\begin{abstract}
We propose an experiment to test Bell's inequality violation in condensed-matter 
physics. We show how to generate, manipulate and detect entangled states 
using ballistic electrons in Coulomb-coupled semiconductor quantum wires. 
Due to its simplicity (only five gates are required to prepare 
entangled states and to test Bell's inequality), the proposed 
semiconductor-based scheme can be implemented with currently 
available technology. Moreover, its basic ingredients may play a role 
towards large-scale quantum-information processing in solid-state devices.
\end{abstract}
\maketitle

The introduction of Quantum Information Processing (QIP) \cite{QIP} has led, 
on the one hand, to unquestionable intellectual progress in understanding 
basic concepts of information/computation theory; on the other hand, 
this has stimulated new thinking about how to realize QIP devices 
able to exploit the additional power provided by quantum mechanics. 
Such novel communication/computation capabilities are primarily related to the 
ability of processing {\it entangled states} \cite{QIP}.
To this end, one should be able to perform precise quantum-state synthesis, 
coherent quantum manipulations ({\it gating}) and detection ({\it measurement}). 
The unavoidable interaction of any realistic quantum system with its 
environment tends to destroy coherence between quantum superpositions. Thus, 
decoherence modifies the above ideal scenario and imposes 
further strong constraints on candidate systems for QIP. Indeed, mainly 
due to the need of low decoherence rates, the only experimental realizations 
of QIP devices originated in atomic physics \cite{AP} and in quantum 
optics \cite{QO}. It is however generally believed that any large-scale application of QIP 
cannot be easily realized with such quantum hardware, which does not allow 
the scalability of existing microelectronics technology. In contrast, 
in spite of the relatively strong decoherence, a solid-state implementation 
of QIP can benefit synergistically from the recent progress in single-electron 
physics \cite{set} as well as in nanostructure fabrication and 
characterization \cite{NANO}.

As already mentioned, the key ingredient for computational speed-up in QIP
is entanglement. Einstein-Podolsky-Rosen (EPR) pairs \cite{epr} 
and three-particle Greenberger-Horne-Zeilinger (GHZ) states \cite{ghz} 
are at the heart of quantum cryptography, teleportation, dense coding, 
entanglement swapping and of many quantum algorithms.
Experimentally, two-particle entangled states have been prepared using 
photons \cite{aspect} and trapped ions \cite{cat_ion}; only recently 
a photonic three-particle entangled state (GHZ) has been also measured
\cite{ghz-photon}.
A few proposals for the generation of entangled states in 
solid-state physics have been recently put forward 
\cite{ddv1}-\cite{barnes}, 
but up to date there are no experimental implementations.

In this Letter we propose an experiment to test 
Bell's inequality violation in condensed-matter physics. 
More specifically, we shall show how to generate, manipulate and 
measure entangled states using ballistic electrons in coupled 
semiconductor quantum waveguides (quantum wires). 
As we shall see, our scheme allows for a direct test of 
Bell's inequality in a solid-state system. To this end, 
a relatively simple gating sequence (five gates only) is identified.

The proposed experimental setup is based on the semiconductor 
quantum hardware of the earlier proposal for {\it quantum computation 
with ballistic electrons} by Ionicioiu {\it et al.}~\cite{quputer}. 
We summarize in the following the main features of this proposal, 
which has been recently analyzed and validated through numerical 
simulations by Bertoni {\it et al.}~\cite{bertoni}.

The main idea is to use ballistic electrons as {\em flying qubits} in 
semiconductor quantum wires (QWRs). In view of the nanometric 
carrier confinement reached by current fabrication 
technology \cite{NANO}, state-of-the-art QWRs behave as 
quasi one-dimensional (1D) electron waveguides. Due to the relatively 
large intersubband energy splittings as well as to the good quality 
of semiconductor/semiconductor interfaces, electrons within the 
lowest QWR subband at low temperature may experience extremely 
high mobility. In such conditions their coherence length can reach 
values of a few microns; therefore, on the nanometric scale 
electrons are in the so-called {\it ballistic regime} and the phase 
coherence of their wave functions is preserved.
This coherent-transport regime is fully compatible with existing 
semiconductor nanotechnology \cite{NANO} and has been the natural 
arena for a number of interferometric experiments with ballistic electrons
\cite{yacoby},\cite{schuster}.
Such fully coherent regime is the basic prerequisite for any QIP.

The building block of our quantum hardware is a pair of adjacent QWR 
structures. The qubit state is defined according to the 
quantum-mechanical state of the electron across this two-wire system.
More precisely, we shall use the so-called {\it dual-rail} representation 
for the qubit: we define the basis state $\ket{0}$ 
by the presence of the electron in one of the wires (called the {\bf 0}-rail) 
and the basis state $\ket{1}$ by the presence of the electron in the other 
one (the {\bf 1}-rail). Saying that the electron is in a given wire 
we mean that: (i) its wave-function is localized on that QWR and 
(ii) its free motion along the wire is well described in terms of a 
quasi-monoenergetic wave-packet within the lowest QWR electron subband 
(with central kinetic energy $E$ and central wave-vector 
$k = \sqrt{2m^* E}/\hbar$).

An appealing feature of the proposed scheme is the mobile character of our 
qubits: using flying qubits we can transfer entanglement from one place 
to another, without the need to interconvert stationary into mobile qubits. 
In the case of stationary qubits (e.g. electron spins in quantum dots) this
is not easily done.

Any QIP device can be built using only single- and two-qubit gates 
\cite{gates}. We choose the following set of universal quantum gates: 
$\left\{ H, P_\varphi, P^C_\pi \right\}$, where 
$H=\frac{1}{\sqrt{2}}\pmatrix{1&1 \cr 1&-1}$ is a Hadamard gate, 
$P_\varphi = \mbox{diag}\,(1,\, e^{i\varphi})$ is a single-qubit phase 
shift, and $P^C_\pi$ is a controlled sign flip. We shall use the more 
general two-qubit gate 
$P^C_\varphi = \mbox{diag}\, (1,\,1,\,1,\,e^{i\varphi})$. 

We now briefly describe the physical implementation of the universal 
quantum gates in terms of the previously introduced dual-rail 
representation. The Hadamard gate can be implemented using an 
{\it electronic beam-splitter}, also called {\it waveguide coupler} 
\cite{alamo,tsukada,resonant_ws}. 
The idea is to design the two-wire system in such a way to spatially 
control the inter-wire electron tunneling. For a given inter-wire distance, a 
proper modulation (along the QWR direction) of the inter-wire potential 
barrier can produce a linear superposition of the basis states 
$\ket{0}$ and $\ket{1}$. More specifically, 
let us consider a {\it coupling window}, i.e.~a 
tunneling-active region, of length $L_c$ characterized by an inter-wire 
tunneling rate $\omega = {2\pi/ \tau}$. As it propagates, the 
electron wave-packet oscillates back and forth between the two 
waveguides with frequency $\omega$. Let $v = {\hbar k/ m^*}$ 
be the group velocity of the electron wave-packet 
along the wire; then, the state $\ket{0}$ goes into the superposition 
$\cos \alpha \ket{0} + \sin \alpha \ket{1}$ with 
$\alpha = \omega t = 2\pi \, {L \over v \tau}$. Let $L_t$ be the 
length necessary for the complete transfer of the electron 
from one wire to the other, 
$\alpha = \pi,\ L_t = v \tau/2$. For a transfer 
length $L_c = L_t/2$ the device is equivalent to a beam-splitter 
and hence, up to a phase shift, to a Hadamard gate. 
By a proper modulation of the inter-wire potential 
barrier we can vary the tunneling rate $\omega$ and therefore the 
rotation angle $\alpha$.
As a result, this structure can operate as a {\sf NOT} gate by adjusting the 
inter-wire potential barrier such that $L_c=L_t$ 
($\pi$-rotation). Similarly, the gate can be turned off by an appropriate 
potential barrier for which the electron wave-packet undergoes a 
full oscillation period, returning back to its original state
($L_c = 2L_t$, $2\pi$-rotation). 
Another way of turning the $H$ gate off is to suppress inter-wire tunneling
by applying a strong potential bias to the coupled QWR structure.

The phase shifter $P_\varphi$ can be implemented using either a 
potential step (with height smaller than the electron energy $V< E$) 
or a potential well along the wire direction; the 
well is preferred since the phase-shift induced is more stable 
under voltage fluctuations. In order to have no reflection from the 
potential barrier, the width $L$ of the barrier should be an integer 
multiple of the half wavelength of the electron in the step/well region, 
$L=n \lambda/2,\ n \in \N$.

\begin{figure}
\putfig{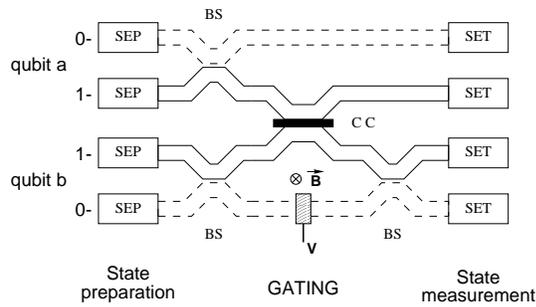}{7}
\caption{Experimental setup to test the Bell-CHSH inequality; 
the ${\bf 0}$-rails of each qubit are dashed for clarity. A potential 
$V$ applied on top of the ${\bf 0}$-rail (dashed box in the figure) 
is used to produce a phase shift $P_{-\theta}^{\bf 0}$ on the second qubit; 
alternatively, the same effect can be achieved with a magnetic field 
$\vec{\mathrm B}$ (via the Aharonov-Bohm effect).} 
\label{bell3}
\end{figure}

We finally describe the two-qubit gate. 
In our scheme the controlled phase shifter $P^C_\varphi$ is implemented 
using a {\it Coulomb coupler} \cite{c_coupler}. 
This quantum gate exploits the Coulomb interaction between 
two single electrons in different QWR pairs (representing the two qubits). 
The gate is similar in construction to the beam splitter previously 
introduced. In this case the multi-wire structure (see Fig.~\ref{bell3}) 
needs to be tailored in such a way 
(i) to obtain a significant Coulomb coupling between the two {\bf 1}-rails 
only and
(ii) to prevent any single-particle inter-wire tunneling. 
Therefore, only if both qubits are in the $\ket{1}$ state
they both experience a phase shift induced by the two-body Coulomb 
interaction. In contrast, if at least one qubit is in 
the $\ket{0}$ state, then nothing happens.

The proposed quantum hardware has some advantages. 
Firstly, the QIP device needs not to be ``programmed'' at 
the hardware level (by burning off the gates), as it may appear. 
Programming is done by switching on/off the gates and this way any 
quantum algorithm can be implemented \cite{applied-fields}. 
Secondly, we use {\em cold programming}, i.e., we set all the gates before 
``launching'' the electrons, so we do not need ultrafast (i.e. 
subdecoherent) electronics for gate operations. 
This property is essential and is a distinct advantage of the proposed 
quantum architecture over other solid-state 
proposals \cite{stationary-qubits}. 
Therefore, the gating sequence needed for the proposed experiment 
can be pre-programmed using {\it static electric fields only}.

One important requirement of our quantum hardware is that electrons within 
different wires need to be synchronized at all times in order to 
properly perform two-qubit gating (the two electron wave-packets should 
reach simultaneously the Coulomb-coupling window).
It is thus essential to have highly monoenergetic electrons
launched simultaneously. This can be accomplished by 
properly tailored energy filters and synchronized single-electron 
injectors at the preparation stage.

\begin{figure} 
\putfig{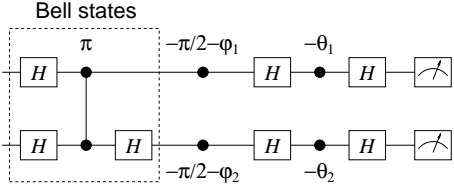}{6}
\caption{Quantum network for the measurement of Bell's inequality. Bell 
states are prepared in the dashed boxed; then the first qubit is measured 
along the direction ${\bf a}=(\theta_1,\varphi_1)$ and the second qubit 
along the direction ${\bf b}=(\theta_2,\varphi_2)$.} 
\label{epr} 
\end{figure}

We now turn to the proposed experimental setup for testing 
Bell's inequality. Two-particle entangled states (Bell states) can 
be generated using three Hadamard gates and a 
controlled-sign shift (see dashed box in Fig.~\ref{epr}; 
the controlled sign shift plus the lower two Hadamards form 
a {\sf CNOT} gate). 
Consider the correlation function for two (pseudo)spins 
$P({\bf a,b})= \langle \sigma^{(1)}_{\bf a} \sigma^{(2)}_{\bf b} 
\rangle$
(here, $\sigma_{\bf a}= \sigma_i a_i$ is the pseudo-spin projection 
along the unit vector ${\bf a}$) \cite{spin}. 
Any local, realistic hidden-variable theory obeys the Bell-CHSH 
\cite{bell,chsh} inequality:
\be 
|P({\bf a,b})+ P({\bf a',b})+ P({\bf a',b'})- P({\bf a,b'})| \le 2 
\label{bell} 
\ee 
This inequality is violated in quantum mechanics. For the singlet 
$\ket{\Psi^-}$, 
a standard calculation gives the result 
\be 
P({\bf a,b})\equiv \langle \Psi^- \vert \sigma_{\bf a}^{(1)} \sigma_{\bf 
b}^{(2)} \vert \Psi^- \rangle = -{\bf a.b} 
\label{ab} 
\ee 
Choosing ${\bf a.b=b.a'=a'.b'=-b'.a}=\sqrt2 /2$, we obtain $2\sqrt{2} \le 2$, 
violating thus Bell inequality (\ref{bell}).

Let us now focus on the correlation function $P({\bf a,b})$. In
the EPR-Bohm gedankenexperiment we need to measure the spin 
component of one particle along a direction $\bf n$. 
However, in our setup this is not possible directly, since we can 
measure only $\sigma_z$, i.e., whether the electron is in the {\bf 0}- or 
in the {\bf 1}-rail. The solution is to do a unitary transformation 
$\ket{\psi} \ra \ket{\psi'}= U \ket{\psi}$, 
such that the operator $\bf \sigma_{\bf n}$ is diagonalized 
to $\sigma_z$, $\bra{\psi} \sigma_{\bf n} \ket{\psi} = \bra{\psi'} \sigma_z \ket{\psi'}$. 
We are looking for a unitary transformation 
$U$ which satisfies $U^+ \sigma_z U= \sigma_{\bf n}$, 
with ${\bf n}= (\sin\theta \cos \varphi, \sin\theta \sin\varphi, 
\cos\theta)$ a unit vector. 
In terms of our elementary gates we obtain
$U(\theta,\varphi)= H P_{-\theta}H P_{-\varphi-\pi/2}$.

Thus, measuring the spin (in the EPR-Bohm setup) along a direction 
${\bf n}$ is equivalent to performing the unitary transformation 
$U(\theta,\varphi)$ followed by a measurement of $\sigma_z$. 
Going back to our entangled pair, we now apply 
on each qubit a local transformation 
$U(\theta_1,\varphi_1)$ and $U(\theta_2,\varphi_2)$, 
respectively. Here, ${\bf a}=(\theta_1,\varphi_1)$ and 
${\bf b}=(\theta_2,\varphi_2)$ are the two directions discussed 
above; at the very end, we measure $\sigma_z$ (i.e., electron in {\bf 0}- or 
{\bf 1}-rail; see Fig.\ref{epr}).

\begin{figure} 
\putfig{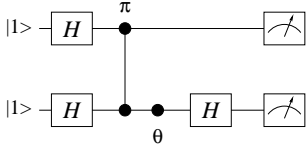}{4}
\caption{Testing Bell's inequality for the singlet state $\ket{\Psi^-}$. 
The quantum network is obtained from Fig.\ref{epr} by setting 
$\varphi_1= \varphi_2= -\pi/2$, $\theta_1=0$ and relabelling 
$\theta=-\theta_2$.} 
\label{epr3}
\end{figure}

For the singlet $\ket{\Psi^-}$ the correlation function depends 
only on the scalar product of the two directions (\ref{ab}), and hence 
only on the angle between them. Without loss of generality we can choose
$\varphi_1= \varphi_2 = -\pi/2,\ \theta_1=0$ and relabel 
$\theta=-\theta_2$. Since $H^2=\bbbone$, the gating sequence 
simplifies to only five gates, as shown in Fig.~\ref{epr3}. 
With this simple network we can measure the correlation function (\ref{ab}) which violates Bell's inequality (\ref{bell}). To perform an Aspect-type experiment \cite{aspect}, we have to choose independently the directions of measurement for each qubit after the electrons are entangled. In this case we need three more gates (after the $P^C_\pi$ gate) $HP_{-\theta_1} H$ on the upper qubit in Fig.~\ref{epr3}.

In practice the situation is more complex. The essential 
ingredient for producing entanglement is the controlled-sign 
shift gate $P^C_\pi$ which involves an interaction between the two qubits. 
Experimentally this requires a good timing of the two electrons 
(they should reach simultaneously the two-qubit gating region).
Suppose that instead of having an ideal $P^C_\pi$ gate 
preparing an ideal singlet (dashed box in Fig.~\ref{epr}), 
in practice we realize a $P^C_\alpha$ gate (possible with unknown 
phase $\alpha$). In this case, instead of preparing the 
singlet $\ket{\Psi^-}$, we end up with the following state:
\be 
\ket{\Psi_\alpha}= \ket{\Psi^-} + e^{i\alpha/2} \cos 
\frac{\alpha}{2} \frac{\ket{1}(\ket{0} -\ket{1})}{\sqrt{2}} 
\ee 
which is a superposition of the singlet and of a separable 
state. 
 
\begin{figure}
\putfig{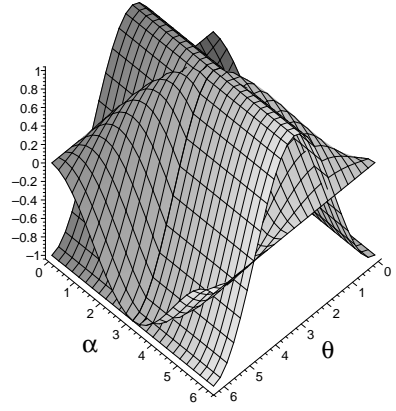}{5}
\caption{Correlation functions $S(\theta)$ and $S(\alpha,\theta)$ 
for the ``ideal'' and ``realistic'' singlet, respectively; 
note that $S(\pi,\theta)=S(\theta)$.} 
\label{epr_cor} 
\end{figure}

Let us now consider the experimental setup discussed above, with 
${\bf a}=(0,\sin\theta_1, \cos\theta_1)$ and 
${\bf b}=(0,\sin\theta_2, \cos\theta_2)$, both in the $Oyz$ plane.
For the correlation function of the imperfect singlet 
$\ket{\Psi_\alpha}$ we obtain 
\be 
S(\alpha,\theta)\equiv \bra{\Psi_\alpha}\sigma^{(1)}_{\bf a}
\sigma^{(2)}_{\bf b} \ket{\Psi_\alpha}= -\sin \frac{\alpha}{2} 
\sin\left(\theta +\frac{\alpha}{2}\right) 
\label{s_alpha} 
\ee
with $\theta=\theta_2-\theta_1$. For $\alpha=\pi$ we recover the correlation function of the singlet,
$S(\theta) \equiv S(\pi,\theta) = -\cos \theta$. 
The two functions are plotted in Fig.\ref{epr_cor}; $S(\theta)$ can be 
identified by noting that there is no $\alpha$ dependence. 

Experimentally, since the one-qubit gate $P_\theta$ is easier to control, 
we can measure the coupling $\alpha$ of the Coulomb coupler $P^C_\alpha$ 
by measuring the dependence of the correlation function 
$S(\alpha,\theta)$ on the phase shift $\theta$ (which can be accurately 
determined). This procedure can be used to determine the purity of the 
singlet, and hence to test/calibrate the Coulomb coupler.

We are now interested to see how small the coupling $\alpha$ can be 
in order to still have a violation of Bell's inequality. The question we ask
is: {\em For what values of $\alpha$ the correlation function 
$S(\alpha,\theta)$ in eq.(\ref{s_alpha}) violates Bell's inequality 
in (\ref{bell})?} 
To this end, we have found a numerical solution: 
the inequality (\ref{bell}) is violated for 
$\alpha \in (\pi/2,\, 3\pi/2)$.
 
A schematic representation of the proposed experimental setup for measuring
Bell's inequality violation is presented in Fig.\ref{bell3}. It is 
possible to reduce the number of gates on the ${\bf 1}$-rail 
by using a phase shifter on the ${\bf 0}$-rail 
$P_{-\theta}^{\bf 0}\equiv \mbox{diag}(e^{-i\theta},\, 1)$
instead of the ${\bf 1}$-rail one used so far
$P_\theta^{\bf 1}\equiv \mbox{diag}(1,\, e^{i\theta})$,
since the two are equivalent (up to an overall phase) 
$P_\theta^{\bf 1}=e^{i\theta} P_{-\theta}^{\bf 0}$.
 
In our setup there are two different ways of producing a phase shift 
$P_\theta$: (i) electrically, with a potential applied on top of the 
${\bf 0}$-rail (the quantum well described above); (ii) magnetically, 
via the Aharonov-Bohm effect, by applying locally a magnetic field 
on the area between the lower two beam-splitters (this can be done since the 
$P_\theta$ and $P^C_\alpha$ gates commute). The second method has the 
advantage of avoiding the no-reflection condition for the potential well 
(the length of the gate should be a half integer multiple of the electron 
wavelength). Either way can be used experimentally.

We stress that Aharonov-Bohm rings and quantum interference 
experiments with ballistic electrons are standard tools in mesoscopic 
physics. A two-slit experiment with an Aharonov-Bohm ring 
having a quantum dot embedded in one 
arm has been reported in \cite{yacoby},\cite{schuster}. This experiment is 
similar to the layout of the lower qubit in Fig.\ref{bell3}, but the 
authors do not use beam splitters and Coulomb couplers. 
In the experimental setup presented here the more difficult part will be to 
implement the Coulomb coupler (CC) and to perform the experiment at the 
single electron level. In our case preparation and measurement 
of the states are done using single electron pumps (SEPs) and single electron 
transistors (SETs) \cite{measure-set}, respectively. 

In conclusion, we have proposed the first measurement of Bell's inequality 
violation in coupled semiconductor nanostructures using ballistic electrons. 
Due to the relative simplicity of the proposed experimental setup
(only five gates are needed to produce entanglement and to test Bell's 
inequality) this measurement scheme is potentially feasible in terms of 
current semiconductor nanotechnology.

R.I.~is grateful to G.~Amaratunga, F.~Udrea and A.~Popescu for stimulating discussions 
and financial support.


\begin{thebibliography}{}
\bibitem{applied-fields}
The phase shifter $P_\varphi$ can be switch on/off by simply switching on/off 
the electric bias applied to the gate. Since $P^C_{2\pi}=\bbbone$, 
this gate can also be turned on/off by means of external biases and 
appropriate design.
\bibitem{spin}
We stress that we are dealing with charge degrees of freedom and the 
present pseudo-spin notation has nothing to do with the physical spin 
of the electrons.
\end{thebibliography}
\end{document}